\documentstyle[prl,aps,preprint]{revtex}
\begin{document}
\newcommand{\kbar}{\shortstack{\hspace*{0.02em}\_\\*[-1.0ex]\it k}}
\draft
\title{Quantum slow motion}
\author{M. Hug and G. J. Milburn}

\address{ 
  Centre for Laser Science,
    Department of Physics,
      The University of Queensland,
        St. Lucia, Qld 4072,
	  Australia
	    }
\date{\today}

\maketitle
\begin{abstract}
We simulate the center of mass motion of cold atoms in a standing, amplitude
modulated, laser field as an example of a system that has a classical mixed
phase-space. We show a simple model to explain the momentum
distribution of the atoms taken after any distinct number of modulation cycles.
The peaks corresponding to a classical resonance move towards 
smaller velocities in comparison to the velocities of the classical 
resonances.
We explain this by showing that, for a wave packet on the classical resonances,
we can replace the complicated dynamics in the quantum
Liouville equation in phase-space by the classical dynamics in a modified
potential. Therefore we can describe the quantum mechanical motion of a 
wave packet on a classical resonance by a purely classical motion.
\end{abstract}
\narrowtext

To have an intuitive picture of the quantum mechanical dynamics of a 
wave packet we are usually confined to the semi-classical regime, that is, to 
orbits with action large compared to Planck's constant 
\cite{gutzwiller,berry77}, or to special systems like the harmonic oscillator, 
where the quantum evolution equations in phase-space are identical to the 
classical ones \cite{hillery}.
In this letter we propose a scheme which enables us to describe a wave packet, 
localized near a resonance of a classical mixed phase-space, by classical 
dynamics in a modified potential.
Hereby we replace the potential in the high order quantum Liouville equation
by an effective potential in such a way that we obtain
a classical Liouville equation. 
Then we describe the quantum motion as classical motion in this modified
potential. We are then able to
characterize the quantum effect by comparing the modified dynamics with the 
dynamics in the original potential. 
This method is applicable well beyond the semi-classical regime for many 
different potentials.

Usually quantum effects on wave packets express themselves in the revival and 
fractional revival properties 
\cite{leichtle} 
or in the occurrence of tunneling phenomena 
\cite{grifoni}.
Both take place on a comparatively long time scale so that we intuitively 
don't expect quantum effects to be visible on a short time scale.
We disprove this intuitive assumption in our model where we use the center 
of mass motion of cold atoms in a standing amplitude modulated laser field.
Here we demonstrate that the momentum distribution
after each cycle of the modulation is peaked at smaller momenta than we would 
expect classically. This shows that the atoms are
traveling slower than we would expect from classical simulations and we can 
give a very simple explanation of this ``quantum slow motion'' phenomenon.  
Since we here include stimulated and spontaneous transitions we expect that 
this quantum mechanical effect is of realistic order of magnitude to observe 
experimentally.  

We investigate a cloud of two level atoms situated in a standing laser field,
with a periodic modulated amplitude.
In this system the Hamiltonian of the center-of-mass motion in the limit of 
large detuning is 
\cite{milburn95}
\begin{equation}\label{eins}
H(t)=\frac{p^2}{2} - \kappa(1-2\epsilon \cos t)\cos q,
\end{equation}
where $p$ and $q$ denote scaled dimensionless momentum and position, $t$ time,
and $\kappa$ and $\epsilon$ are the parameters defining the depth of the
standing wave and the strength of the amplitude modulation, respectively.
Note that $p$ and $q$ fulfill the commutator relation
$[p,q]={\rm i} \kbar$,
where $\kbar$ is a scaled Planck's constant that is in some sense a measure 
for the ``quantum mechanicality'' of the problem since it defines the size of a 
minimum uncertainty wave packet in relationship to the resonances
\cite{graham}.

In Fig. 1(left) we show as an example the classical stroboscopic
phase-space portrait 
\cite{milburn93} 
for $\epsilon=0.2$  
with $\kappa=1.2$. This choice of parameters is capable to show
classical stable period-one resonances after each modulation period  
symmetrically situated along the  momentum axes.
This specific phase-space structure allows a quantum mechanical wave packet, 
situated initially near one of these resonances, to coherently  
tunnel to the other resonance. 
This takes place on a long time scale  in terms of cycles of the modulation. 
Note that this tunneling cannot be understood in terms of the  
presence of a potential barrier as it is present in several publications 
\cite{grifoni}
regarding tunneling in mixed systems.

We simulate the tunneling dynamics by starting each realization with a minimum 
uncertainty wave packet that may be squeezed 
\cite{}, 
centered on the classical
resonance. We then simulate the full quantum mechanical 
dynamics by applying a split operator algorithm  with adapted
time step size \cite{feit}
in the context of a standard quantum
Monte Carlo integration scheme to include stimulated and spontaneous
transitions 
\cite{milburn95}.
We calculate  the mean momenta and the corresponding variance from the 
Poincare section of the
momentum distribution taken after each cycle of the modulation at $t=2n\pi$.
In Fig. 2 (full line) we show the result of this simulation for 
$\epsilon = 0.2$, $\kappa=1.2$, and $\kbar=0.25$. 
Related to recent experiments 
\cite{acht}
we used the parameters for Rubidium to obtain a realistic scenario.
We plot the mean momentum after each cycle of the 
modulation of the standing wave against the number of cycles. As expected
and clearly indicated by the drop down of the variance, 
we observe coherent tunneling of  the mean momentum from the location of the 
resonance at approximately $p=1$ to the corresponding resonance at $p=-1$. 

However there are additional oscillations that might lead to the conclusion 
that the wave packet is not sitting precisely on the classical period-one 
fixed point but is indeed circulating around an alternative stable point in 
phase-space. 
It seems like the wave packet, centered on the classical resonance  
is not appropriately centered on the "true" resonance but sitting beside it. 
Therefore the mean momentum at each kick strongly oscillates around its mean 
motion.

This lead us to the conclusion that if we moved the initial wave packet onto 
this alternative stable point and started the simulation of the dynamics from 
there, we expect the oscillations to vanish. 
This is exactly what we see in Fig. 2 (dashed line).
The oscillations are strongly compressed and we face essentially the
situation of a well localized wave packet which undergoes coherent tunneling on
the longer time scale. For the dynamics of the wave packet the classical 
resonance is obviously not important but a modified resonance, shifted 
towards slower momentum. How can we explain this effect?

To give an explanation we first recall that a wave packet localized near a 
classical resonance has been shown 
\cite{milburn93} 
to remain localized without 
changing its shape, at least for a long time. 
Therefore we may assume that a minimum uncertainty wave packet sitting near 
a classical resonance will  remain unchanged in shape for several cycles. 
This is the main assumption we need to apply a theory of 
Henriksen et. al 
\cite{henriksen} 
where the effect of quantum mechanics on a wave packet is described as 
classical motion, that is as motion following the classical Liouville 
equations in phase-space, but in a modified potential.

The convenient quantum mechanical phase-space representation is the Wigner 
function $W(q,p,t)$,because it has the correct quantum mechanical  marginal
distributions. 
Since in the experiments we are seeking to  describe the momentum distribution 
and the position distribution of the center of mass motion, this property of 
the Wigner function allows us to compare the marginals directly with the 
measured distributions.

The phase-space dynamics of the Wigner function is given by 
\cite{wigner,henriksen}
\widetext
\begin{equation}\label{wigner}
\frac{\partial W}{\partial t}=-p\frac{\partial W}{\partial q}
                +\frac{\rm i}{\kbar}
		\left(
		\sum\limits_{\nu=0}^{\infty} \frac{1}{\nu!}
		    \left(\frac{\kbar}{2{\rm i}}\right)^{\nu}
		          \frac{\partial^{\nu}V(q,t)}{\partial q^{\nu}}
			  \frac{\partial^{\nu}W}{\partial p^{\nu}}
	        -          
		\sum\limits_{\nu=0}^{\infty} \frac{1}{\nu!}
		    \left(-\frac{\kbar}{2{\rm i}}\right)^{\nu}
		           \frac{\partial^{\nu}V(q,t)}{\partial q^{\nu}}
			   \frac{\partial^{\nu}W}{\partial p^{\nu}}
	        \right)
\end{equation}
where $V(q,t)=\kappa(1-\epsilon \cos t) \cos q$ denotes the potential.
This, for the following convenient representation, corresponds to the well 
known one given by Wigner 
where  only one sum over odd 
derivatives occurs. 

Due to the special spatial dependence of our cosine potential, where the odd 
derivatives reproduce themselves, we can replace the infinite sum by defining 
an effective potential $V_{eff}$ by
\begin{equation}
\frac{\partial V_{eff}}{\partial q}=
      \frac{\rm i}{\kbar}
      \left(
      \sum\limits_{\nu=0}^{\infty} \frac{1}{\nu!}
          \left(\frac{\kbar}{2{\rm i}}\right)^{\nu}
          \frac{\partial^{\nu}V(q,t)}{\partial q^{\nu}}
	  \frac{\partial^{\nu}W}{\partial p^{\nu}}                
      -                         
      \sum\limits_{\nu=0}^{\infty} \frac{1}{\nu!}
          \left(-\frac{\kbar}{2{\rm i}}\right)^{\nu}
	  \frac{\partial^{\nu}V(q,t)}{\partial q^{\nu}}
	  \frac{\partial^{\nu}W}{\partial p^{\nu}}
      \right)/\frac{\partial W}{\partial p}.
\end{equation}
\narrowtext
Then Eq. \ref{wigner} is replaced by the first order equation
\begin{equation}
\frac{\partial W}{\partial t}=
            -p\frac{\partial W}{\partial q}
	    +\frac{\partial V_{eff}}{\partial q}\frac{\partial W}{\partial p}
\end{equation}
which is identical to the  classical Liouville equation describing the classical
dynamics in the modified potential $V_{eff}$. In this sense the action of 
quantum mechanics can be described by the classical motion in a modified 
potential.  

Assuming a Gaussian squeezed minimum uncertainty  wave packet with time dependent 
squeeze parameter $\xi(t)$, we take the Wigner function to be of the form
\begin{equation}
W(q,p,t)=\frac{1}{\pi\kbar}\exp\left(-\frac{\xi}{\kbar}(q-\!\langle q \rangle)^2
          -\frac{1}{\kbar \xi}(p-\!\langle p \rangle)^2\right)
\end{equation}
with the mean time dependent momentum and position, $\langle p \rangle(t)$ and 
$\langle q \rangle(t)$, respectively, chosen in such a way, that the wave packet 
always stays  centered on the resonance in order the assumption of staying
unchanged in shape to remain valid. 
It is not important to know the explicit time dependence of these 
parameters.
Then the effective potential is 
\begin{equation}\label{veff}
V_{eff}(q,t)=
     V(q,t) \exp\left(-\frac{\kbar}{4}\right)
            \frac{\sinh (p-\langle p \rangle)}{p-\langle p \rangle}
\end{equation} 
That means the motion of the wave packet is locally described by the original 
potential compressed by a factor of $\exp(-\kbar/4)$ since 
the sinh-factor can, for the sake of qualitative discussion, locally be 
approximated by $1$.

In Fig. 1 (middle and right) we show for $\kbar=0.25$ and $\kbar=0.35$
classical stroboscopic phase-space portraits for the 
effective potential and compare them to the phase-space portrait of the 
original potential. 
Note, that our approximation is only valid in the vicinity 
of the period-one resonances. However, since we are interested in exactly
these regions of phase space  this kind of representation gives an idea of what is
going on, although the other phase-space regions are not to be taken as a valid
description of the dynamics there.
The main conclusion regarding the resonances is that the central resonance 
at $(q,p)=(0,0)$
becomes smaller and the second order resonances we are interested in are 
pushed towards smaller momenta $p$,  which exactly corresponds to 
the observation made in Fig. 2, where we could simulate the tunneling 
phenomenon best for initially situating the wave packet at the shifted
resonance.  

Eq. \ref{veff} indicates that the effect scales with $\kbar$ which 
identifies it as a purely quantum mechanical effect. 
We can clearly see this property by comparing Fig. 1 (middle) and (right), where we can
directly see the relocation of the classical resonance for two values of 
$\kbar$.
In Fig. 3 we simulate wave packets for different values of $\kbar$ for only a 
few cycles.
We start each simulation with a minimum uncertainty wave packet in such a way
that the oscillations in the evolution are most suppressed and observe that
in correspondence to the modified potential the mean momenta and therefore
the wave packets themselves are relocated towards smaller velocities with 
increasing $\kbar$. 
Note that the curve corresponding to $\kbar=0.25$ corresponds to a situation 
where the conditions for tunneling are fulfilled, therefore the mean momentum 
starts to decrease.

There is a second important consequence of this phenomenon in the 
scenario  of present experiments 
\cite{acht} 
of investigating the short time 
behavior of loading all the resonances from a spatially uniform distributed 
cloud of atoms.
In order to effectively load the resonances we start with a phase shift of 
$-\pi/2$,  that is to say, we now investigate the Hamiltonian 
\begin{equation}\label{zwei}
H(t)=\frac{p^2}{2} - \kappa(1-2\epsilon \sin t)\cos q,
\end{equation}
and take the snapshots at $t=\pi/2+2n\pi$.
Then the resonances are initially aligned on the q-axes and are therefore 
covered best by the cloud of atoms.
A classical picture of the dynamics suggests, that
as time goes by only those atoms initially sitting close to a resonance remain,
whereas all the other atoms perform a nonlinear motion corresponding to
the fact that they are sitting in a chaotic region \cite{zurek} of phase-space.
Therefore we expect to observe after some time only the three peaks of
loaded resonances.
Since the assumption of a durable wave packet is only valid for a wave packet 
initially situated on a resonance and not for all the other wave packets this  
motivates us to believe that the local relocation of the resonance described 
above only happens to those atoms trapped at the resonance.
This should change the overall momentum distribution in comparison to a pure 
classical simulation.

In Fig. 4 (left) we compare the momentum distributions of snapshots at
$t=9\pi/2$, that is after 2.25 modulation cycles only, of three
different simulations: a  quantum mechanical (top), a modified classical
(middle), and a purely classical simulation(bottom). Note that this is a very
short time compared to tunneling and revival experiments.
For the quantum simulation (top) we start with a large number of wave packets
of the width of the distribution in momentum of the atom cloud. 
The width in position is chosen in order to have a 
minimum uncertainty wave packet.
We distribute them uniformly on the q-axis, apply the Monte Carlo 
integration scheme and finally add the contribution of each wave packet up 
to get the whole momentum distribution. In the purely classical simulation 
(bottom) 
we simply take a cloud of point particles uniformly distributed 
in $q-$direction and Gaussian distributed in $p-$direction. 
Then the individual motion of the atoms is treated classically by letting 
the atoms evolve following the classical Liouville dynamics, 
but we still have included stimulated and spontaneous transitions in a 
Monte-Carlo-Integration scheme. 

Note that the quantum peaks are shifted towards 
smaller momenta. This shift becomes larger with the scaled Planck's constant 
$\kbar$ which is a further indication, that this effect can  be explained by
the quantum mechanical effect described above.
To show that the occurrence of the effective potential may in principle be
sufficient to explain this feature,
we simulate this by applying the classical simulation again (middle), where we 
now change the trajectory according to the effective potential once we start 
on a resonance.
This is a very simple approach which is certainly only useful to show 
qualitatively that our explanation is suitable to describe the quantum dynamics.
But we note that this modified classical simulation indeed shows the essential 
features of the pure quantum simulations.

In Fig. 4(right) we show the same simulations but without any modulation.
Here the difference corresponding to the quantum mechanical effect vanish and
now more or less all three simulations show the same structure. This structure
is due to classical transient effects, which appear in the first few cycles
and are closely related to the motion in the standing wave, 
since they are independent of the modulation.  This transient is always there 
and interferes with the quantum
mechanical effect investigated in this paper. However the quantum mechanical 
effect is easily to identify since it vanishes for $\epsilon=0$. 
Therefore this effect is clearly related to the modulation and therefore shows 
a quantum feature of the classical mixed phase space. 
Note that the effect is vanishing for $\epsilon=0$  is consistent with our 
theory since in this case we face classical integrable motion. 
A wave packet in such a system is not stabilized but spreads and changes its 
shape and therefore the assumption for applying the theory of Henriksen et al.
is no longer valid.
 
To conclude, we have shown that we can use the property of wave packets
staying localized on resonances of a classical mixed phase space to
simplify the complicated quantum dynamics in phase space. 
In this case we can describe the quantum dynamics of the wave packet by the 
classical motion in a modified potential. This is not only valid for the 
cosine potential investigated, but also as already mentioned in 
\cite{henriksen} to polynomial potentials of arbitrary high order and  
to other systems that has been topic of investigations of the relationship of 
classical chaotic motion and the corresponding quantum dynamics. 
First there is the atomic bouncer in an evanescent field 
\cite{milburn97a}
\begin{equation}
V(q,t)=\lambda q + \kappa(1+\epsilon \cos t)\exp(-q)
\end{equation}
This setup of evanescent light waves can be modified to get a Morse potential 
\cite{ovch} which serves as an atomic trap. In these two cases it is  also very
straightforward to find the modified potential and to come to similar
conclusions to those in this paper.

\begin{figure}
\caption{\label{1} Left: Stroboscopic phase space portrait of the classical 
motion described by the Hamiltonian Eq. \protect\ref{eins} for $\kappa=1.2$ 
and $\epsilon=0.2$. Middle and right:
Stroboscopic phase space portraits of the corresponding effective
potentials Eq. \protect\ref{veff} with $\protect\kbar=0.25$ and $\protect\kbar=0.35$ }
\end{figure}
\begin{figure}
\caption{\label{2} 
Left: Mean momentum $\langle p \rangle$ of the quantum mechanical simulation of the 
dynamics of two wave packets in dependence of number of cycles $s$. The first
(straight line) is initially sitting on the classical resonance $(p_m=1.03)$, the 
second (dashed line) on the modified at $p_m=0.84$. 
Here the parameters are $\protect\kbar=0.25, \kappa=1.2$, and $\epsilon=0.2$
Right: Corresponding variance $V[p]$. }
\end{figure}
\begin{figure}
\caption{\label{3} Mean momentums $\langle p \rangle$ of the quantum mechanical 
simulation of the 
dynamics of several wave packets in dependence on number of cycles $s$
with $\kappa=1.2$ and $\epsilon=0.2$. Here
$\protect\kbar$ takes on the values $0.15,0.2,0.25,0.3$ (from top). }
\end{figure}
\begin{figure}
\caption{\label{4} Left:  Quantum, modified classical, and purely classical 
simulation (from top) of momentum distributions $P[p]$ of snapshots after 2.25 
modulation cycles  of the Hamiltonian Eq.\protect\ref{zwei} with
$\kappa=1.2, \epsilon=0.2$, and $\protect\kbar=0.35$.
Right: The same simulations but for the unmodulated case $\epsilon=0$.
} 
\end{figure}
%\begin{figure}
%\caption{\label{5} Same as Fig. \ref{4}, but now after 4 cycles $(t=3.25)$}
%\end{figure}

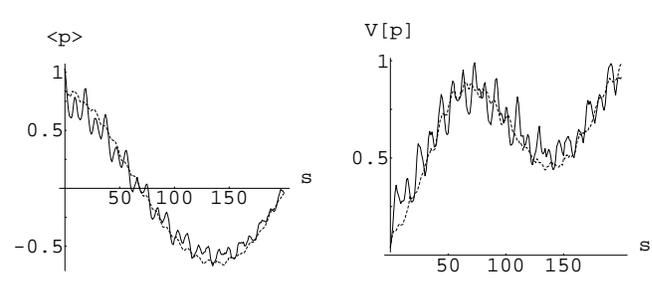
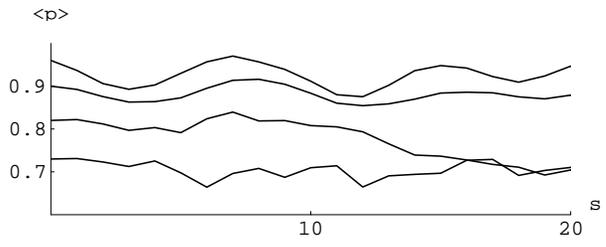
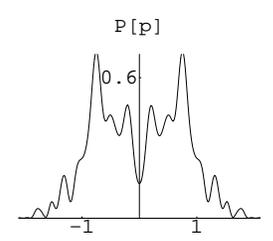
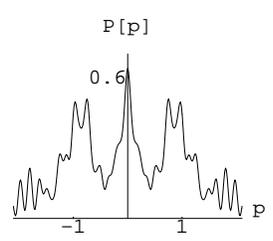
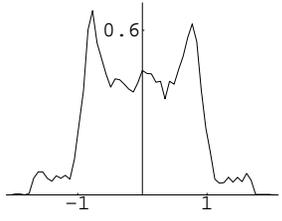
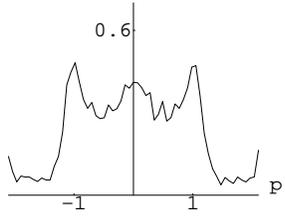
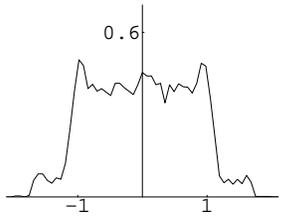
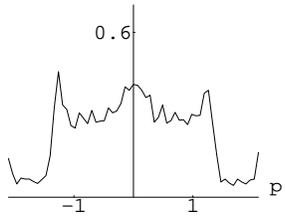
\end{document}